\documentclass[aps,prd,twocolumn,showpacs,superscriptaddress,groupedaddress,nofootinbib,floatfix,longbibliography]{revtex4-1}

\usepackage{amsmath}
\usepackage{amssymb}
\usepackage{graphicx}
\usepackage{hyperref}
\usepackage{color}
\usepackage{multirow}
\usepackage{booktabs}
\usepackage{bbm}
\usepackage{float}
\usepackage{fixltx2e}
\usepackage{tikz}
\usetikzlibrary{arrows.meta, positioning, fit, calc, backgrounds}

\newcommand{\prog}[1]{\texttt{#1}}

\begin{document}

\title{Learning to Unscramble Feynman Loop Integrals with \textsc{SAILIR}}

\author{David Shih}
\email{shih@physics.rutgers.edu}
\affiliation{NHETC, Dept.\ of Physics and Astronomy, Rutgers University, Piscataway, NJ 08854, USA}

\date{\today}

\begin{abstract}
Integration-by-parts (IBP) reduction of Feynman integrals to master integrals is a key computational bottleneck in precision calculations in high-energy physics. Traditional approaches based on the Laporta algorithm require solving large systems of equations, leading to memory consumption that grows rapidly with integral complexity. We present \textsc{Sailir} (\textbf{S}elf-supervised \textbf{AI} for \textbf{L}oop \textbf{I}ntegral \textbf{R}eduction), a new machine learning approach in which a transformer-based classifier guides the reduction of integrals one step at a time in a fully online fashion. The classifier is trained in an entirely self-supervised manner on synthetic data generated by a scramble/unscramble procedure: known reduction identities are applied in reverse to build expressions of increasing complexity, and the classifier learns to undo these steps. When combined with beam search and a highly parallelized, asynchronous, single-episode reduction strategy, \textsc{Sailir} can reduce integrals of arbitrarily high weight with bounded memory. We benchmark \textsc{Sailir} on the two-loop triangle-box topology, comparing against the state-of-the-art IBP reduction code \prog{Kira} across 16 integrals of varying complexity. While \textsc{Sailir} is slower in wall-clock time, its per-worker memory consumption remains approximately flat regardless of integral complexity, in contrast to \prog{Kira} whose memory grows rapidly with complexity. For the most complex integrals considered here, \textsc{Sailir} uses only 40\% of the memory of \prog{Kira} while achieving comparable reduction times.
This demonstrates a fundamentally new paradigm for IBP reduction in which the memory bottleneck of Laporta-based approaches could be entirely  overcome, potentially opening the door to precision calculations that are currently intractable.

\end{abstract}

\maketitle

\section{Introduction}
\label{sec:intro}

The reduction of Feynman integrals to a basis of master integrals via integration-by-parts (IBP) identities~\cite{Chetyrkin:1981qh,Tkachov:1981wb} is an essential step in virtually all state-of-the-art precision calculations in the Standard Model and beyond. As calculations push to higher loop orders and more external legs, IBP reduction has become one of the primary computational bottlenecks, requiring significant time and memory resources (see~\cite{Smirnov:2025dfy} for a recent review).

The standard approach to IBP reduction is the Laporta algorithm~\cite{Laporta:2000dsw}, which generates a large system of linear equations from IBP identities and solves them using Gaussian elimination. Several mature software packages implement variants of this algorithm, including \prog{Kira}~\cite{Maierhofer:2017gsa,Klappert:2020nbg,Lange:2025fba}, \prog{FIRE}~\cite{Smirnov:2019qkx,Smirnov:2025prc}, \prog{Reduze}~\cite{Studerus:2009ye,vonManteuffel:2012np}, and \prog{LiteRed}~\cite{Lee:2013mka}. A key challenge of the Laporta approach is that the system of equations---and hence the memory required---grows rapidly with the complexity of the integrals being reduced. For state-of-the-art multi-loop calculations, memory consumption of tens or hundreds of gigabytes is common, and the memory wall is often the limiting factor.

Several strategies have been developed to mitigate the memory problem within the Laporta framework. Finite-field methods~\cite{Kant:2013vta,vonManteuffel:2014ixa,Peraro:2019svx,Klappert:2019emp} replace exact rational arithmetic with modular arithmetic, dramatically reducing coefficient sizes. \prog{Kira} combines finite-field reconstruction with algebraic simplification to achieve the current state of the art. Improved seeding strategies---choosing which IBP identities to generate so as to minimize the system size---have also yielded significant gains; see e.g.\ \prog{NeatIBP}~\cite{Wu:2023upw} and \prog{Blade}~\cite{Guan:2024byi}.

There has also been considerable development of alternatives to the Laporta algorithm. For example, one strategy is to derive symbolic reduction rules that can be applied iteratively to reduce the weight of an integral, avoiding the need to solve a large linear system~\cite{Lee:2013mka,Kosower:2018obg,Feng:2025leo,Liu:2025udl,delaCruz:2026mas,Smith:2025xes}. This approach does not always succeed in practice, but can be very effective when it does. 

Recent work has also started to explore machine learning (ML) techniques for IBP reduction. In \cite{vonHippel:2025okr,Song:2025ibp,Zeng:2025ibp}, ML methods such as \prog{FunSearch} \cite{2024Natur.625..468R}, genetic programming, and simulated annealing were employed to produce (meta)heuristics and priority functions for improving seed selection  within the Laporta framework. Ref.~\cite{Zeng:2025ibp} also considered the use of reinforcement learning (RL) for IBP reduction: instead of the Laporta algorithm, the goal was to train an agent to select IBP actions step by step. The RL-based approach was demonstrated for two integrals taken from the topology of one-loop bubble integrals.

In this paper, we develop a new machine-learning approach to IBP reduction. Like \cite{Zeng:2025ibp}, we also view IBP reduction as a sequential decision problem -- a Markov decision process (MDP) -- where an agent is trained to predict the best IBP identity to apply to the expression at every step.  By turning IBP reduction into a fully online process, we avoid having to store giant linear systems in memory. In principle  this would allow  the memory limitation  of  Laporta-based approaches to be entirely overcome, provided the MDP was highly performant and able to successfully find the correct actions to successfully reduce integrals of any weight. Providing a highly performant MDP approach to IBP reduction is the aim and central result of this work. 

Our approach, which we call \textsc{Sailir} (\textbf{S}elf-supervised \textbf{AI} for \textbf{L}oop \textbf{I}ntegral \textbf{R}eduction), combines three main ideas:

\begin{itemize}

\item \emph{Self-supervised training via unscrambling.}  Instead of training it with RL, which is very computationally expensive and struggles with a sparse reward landscape, we instead train it with labeled reduction trajectories generated without requiring an existing IBP solver, using the scramble-and-unscramble framework of Ref.~\cite{Shih:2026lmy}. In that work, self-supervised oracle trajectories were used to train ML models for simplifying mathematical expressions (dilogarithm identities and scattering amplitudes), where the starting expression is complex and the goal is a simpler one. IBP reduction presents a qualitatively different challenge: the goal is to reduce the \emph{weight} of integrals---a measure of integral complexity defined precisely in Eq.~\eqref{eq:weight}---starting from a single high-weight integral and ending with a longer expression involving lower-weight integrals. We adapt the scramble-and-unscramble framework to this setting by exploiting the linearity of IBP identities to reorder the unscramble, targeting integrals by weight rather than in scramble order, which produces training data better aligned with the inference task.

\item \emph{Learning-to-rank action classifier.} The number of valid IBP identities at each step varies from tens to thousands, making a fixed-output-size architecture impractical. We adopt a poly-encoder architecture from the learning-to-rank literature~\cite{Humeau:2020polyencoders}, where the expression state is encoded once and used to score each candidate action independently. This allows efficient inference over variable-sized action spaces.

\item \emph{Hierarchical parallel reduction with bounded memory.} Each integral is reduced independently by a worker process that reduces the maximum weight by one level (we refer to this as a single \emph{episode} of the reduction), using beam search guided by the trained classifier. Since reducing one integral typically introduces new integrals that themselves need reduction, we employ an asynchronous orchestrator that submits workers as needed, caches results, and reuses them when the same integral appears in multiple reduction paths. Crucially, each worker's memory consumption is bounded and independent of the complexity of the integral being reduced.
\end{itemize}

Notably, \textsc{Sailir} only requires training once per integral family (topology): the model is trained on scrambled trajectories sampled across all sectors and weights of the family, and then it becomes capable of reducing any integral at any weight from that family. 

Following standard practice in modern IBP reduction~\cite{Kant:2013vta,vonManteuffel:2014ixa,Peraro:2019svx,Klappert:2019emp}, we work with numerical kinematics and finite-field arithmetic throughout, which keeps all coefficients as single machine-precision integers and eliminates the need for exact rational arithmetic.

We benchmark \textsc{Sailir} on the two-loop triangle-box topology studied in Ref.~\cite{vonHippel:2025okr}, comparing against \prog{Kira} across 16 integrals covering a broad range of weights. Our main findings are:
\begin{itemize}
\item All 16 integrals are successfully reduced to the known set of 16 master integrals, with a 100\% success rate.
\item Per-worker memory remains approximately flat at ${\sim}$3~GB across all complexities, while \prog{Kira}'s memory grows from 159~MB at the lowest weights to 8.7~GB at the highest weights.
\item For the most complex integrals, \textsc{Sailir} achieves memory ratios of $0.4$--$0.7\times$ compared to \prog{Kira}, with time ratios converging to $1.0$--$4.7\times$.
\end{itemize}

The remainder of this paper is organized as follows. Section~\ref{sec:ibp_background} reviews the IBP reduction framework and introduces notation. Section~\ref{sec:framework} presents the general framework of \textsc{Sailir}. Section~\ref{sec:triangle_box} demonstrates the framework on the two-loop triangle-box topology, showing that the flat memory consumption of \textsc{Sailir} eventually wins out over \prog{Kira}'s rapidly rising memory consumption for sufficiently high-weight integrals. Section~\ref{sec:discussion} contains a brief summary and discussion of future directions. Various additional technical details are included in the appendices.

\section{IBP Reduction Background}
\label{sec:ibp_background}

An $L$-loop Feynman integral family is defined by a set of $N_p$ propagators $D_{i}$, $i=1,\dots,N_p$ and $N_s$ irreducible scalar products (ISPs) $D_{i}$, $i=N_p+1,\dots,N_p+N_s$. These form a complete basis for the quadratic Lorentz invariants built out of the external and loop momenta. (Invariants that involve only external momenta are not relevant for loop integral reduction and are excluded from this counting.) They appear in the integrand as
\begin{equation}
I[\mathbf{a}]=\int \prod_{j=1}^{L} d^d k_j \prod_{i=1}^{N_p+N_s}{1\over D_i^{a_i}}
\end{equation}
A general integral in the family is specified by an index vector $\mathbf{a} = (a_1, \ldots, a_{N_p + N_s})$, where $a_i \geq 1$ indicates present in the denominator while $a_i<0$ indicates present in the numerator. Propagators can appear in the denominators of integrands, while ISPs do not correspond to any propagator and can only appear in numerators of integrands.

 The integral family is organized into $2^{N_p} - 1$ non-trivial \emph{sectors}, each defined by the subset of propagators with positive power (i.e.\ present in the denominator). A sector $S'$ is a {\it subsector} of $S$ if all of its denominator factors are present in $S$.

Integration-by-parts identities arise from the translational invariance (shift symmetry) of loop momenta in dimensional regularization, which implies the vanishing of total derivatives:
\begin{equation}
\int \prod_{j=1}^{L} d^d k_j \; \frac{\partial}{\partial k_i^\mu} \left( v^\mu \cdot \text{integrand} \right) = 0,
\label{eq:ibp_identity}
\end{equation}
where $i \in \{1, \ldots, L\}$ selects the loop momentum to differentiate with respect to, and $v^\mu$ ranges over all $L + E_{\mathrm{indep}}$ independent momenta (loop momenta $k_1, \ldots, k_L$ and independent external momenta), giving $L(L + E_{\mathrm{indep}})$ IBP equation templates. Each template, evaluated at a specific \emph{seed integral} $I[\mathbf{a}]$, produces a linear relation among integrals with shifted indices. Lorentz invariance (LI) identities provide additional relations: since the integral is a Lorentz scalar, infinitesimal rotations in the plane of two external momenta $p_\mu, p_\nu$ leave it invariant, yielding $\binom{E_{\mathrm{indep}}}{2}$ further equation templates. Together, the IBP and LI identities generate all linear relations among integrals in the family.\footnote{A third class of relations, \emph{symmetry relations}, can further reduce the system by exploiting symmetries of the propagator set under relabelings of loop and external momenta, mapping equivalent sectors onto each other (see e.g.~\cite{Argeri:2007up} for a pedagogical overview). We do not use symmetry relations in this work, but they could be incorporated in future extensions of \textsc{Sailir}.}
  
A key structural property is that IBP and LI identities can never introduce a propagator absent from the seed integral. This follows from the chain rule: differentiating $D_j^{-a_j}$ produces $-a_j \cdot D_j^{-(a_j+1)}$, so every term raising index $a_j$ carries a coefficient proportional to $a_j$. When $a_j = 0$, this vanishes. Consequently, identities evaluated at a seed in sector $S$ produce integrals only in $S$ or its subsectors---never in higher sectors (which contain additional propagators) or lateral sectors (sectors at the same level that are not subsectors of one another). We exploit this property during training data generation (Section~\ref{sec:framework}): scrambling with identities seeded within a sector automatically preserves the sector hierarchy, requiring no explicit sector filtering.

The goal of IBP reduction is to express any integral in the family as a linear combination of a minimal set of \emph{master integrals}---integrals that cannot be further reduced by any IBP or LI identity. The number of masters and their identity are determined by the topology and kinematics. The \emph{corner integral} of a sector is the simplest integral with all propagators at unit power and no ISP insertions. The set of all corner integrals forms an overcomplete basis: the true set of master integrals is generally a subset. However, reducing an arbitrary integral down to corner integrals constitutes the bulk of the computational challenge, since it requires eliminating all ``dots'' (raised propagator powers) and ISP insertions. Relations among corner integrals themselves are typically much simpler. In this work we train \textsc{Sailir} on the reduction to corner integrals, but find that the trained model generalizes for free to the full reduction to master integrals (Section~\ref{sec:results}).
 
To facilitate IBP reduction, it is convenient to define a total order on integrals within a given sector by assigning each a weight. We use the lexicographic tuple
\begin{equation}
w(I[\mathbf{a}]) = \Big(\sum_{i} \max(a_i, 0),\;\; \sum_{i} |{\min(a_i, 0)}|,\;\; \mathbf{w}_3\Big),
\label{eq:weight}
\end{equation}
where $r \equiv \sum_i \max(a_i, 0)$ is the total positive-index weight (i.e.\ the sum of all denominator powers, including dots), $s \equiv \sum_i |\min(a_i, 0)|$ is the total numerator power, and $\mathbf{w}_3 = (|a_1|, \ldots, |a_{N_p+N_s}|)$ is a lexicographic tiebreaker over individual indices. Both the Laporta algorithm and \textsc{Sailir} use this weight ordering to organize the reduction: integrals are eliminated in decreasing weight order, with each elimination expressing a higher-weight integral in terms of lower-weight ones. In the Laporta algorithm, this is achieved by generating a large system of IBP equations and solving it via Gaussian elimination ordered by weight. In \textsc{Sailir}, each integral is eliminated individually by applying a sequence of IBP+LI identities chosen by a trained classifier.

\section{\textsc{Sailir}: General Framework}
\label{sec:framework}

\subsection{IBP reduction as an MDP}

With \textsc{Sailir}, we view IBP reduction as a deterministic Markov decision process (MDP). An MDP is a sequential decision-making framework in which an agent interacts with a system through a series of discrete steps. At each step, the agent observes the current \emph{state} of the system, selects an \emph{action} from a set of allowed actions, and the system transitions to a new state determined by the chosen action. In our setting, the state encodes the current expression being reduced, and an action corresponds to applying a specific IBP or LI identity. The agent's goal is to select a sequence of actions that reduces the expression to a combination of lower-weight integrals.

The state at each step consists of four components: (i)~the current expression, a dictionary $\{I[\mathbf{a}]: c\}$ mapping integrals to their coefficients; (ii)~the substitution history, recording previously solved integrals and their replacement expressions; (iii)~the target integral, defined as the highest-weight non-master integral to eliminate; and (iv)~the sector mask, an $N_p$-bit encoding of the current sector.

Each IBP+LI identity is labeled by a template index $a \in \{1, \ldots, N_{\text{op}}\}$, which selects the IBP or LI template, and a seed integral $\mathbf{s} \in \mathbb{Z}^{N_p + N_s}$, at which the template is evaluated. Each template defines a set of shifts $\boldsymbol{\delta} \in \mathcal{S}_a$ relative to the seed, and the identity takes the form
\begin{equation}\label{eq:IBPidentgen}
\sum_{\boldsymbol{\delta}\in \mathcal{S}_a} c_a(\mathbf{s}+\boldsymbol{\delta})\, I[\mathbf{s}+\boldsymbol{\delta}] = 0
\end{equation}

A \emph{valid action} on a target integral $\mathbf{t}$ is a choice of IBP+LI identity $(a, \mathbf{s})$ such that, after applying all accumulated substitutions to Eq.~\eqref{eq:IBPidentgen}, the target $\mathbf{t}$ appears with non-zero coefficient, and no integrals outside the target's subsector are introduced. Applying the action means solving the resulting equation for $\mathbf{t}$:
\begin{equation}
\mathbf{t} = -c_{\mathbf{t}}^{-1} \sum_{\mathbf{a}\ne\mathbf{t}} c_{\mathbf{a}} I[\mathbf{a}] 
\end{equation}
updating the expression and adding $\mathbf{t} \to \text{solution}$ to the substitution history.

Note that as the substitution history grows, it enables {\it indirect} actions as more integrals are solved. These are actions where the target appears only after substitution, and they are a key feature of the MDP.  

To ensure efficient single-pass application, we maintain a \emph{resolved} substitution dictionary -- when a new entry is added, all existing values are updated and the new value is expanded using all prior substitutions. This way no value in the substitution dictionary references another key and the substitutions can be applied with a single pass. Without this, the substitutions would need to be applied recursively, leading to significant computational overhead as the substitution history grows.

\subsection{Training data}
\label{sec:trainingdata}

We generate training data for the MDP using the scramble-and-unscramble
framework of Ref.~\cite{Shih:2026lmy}. In that work, self-supervised
reduction trajectories were generated for training ML models to simplify
mathematical expressions (dilogarithm identities and scattering
amplitudes). The key idea is that starting from a simple
expression, one can \emph{scramble} it by applying a sequence of random
algebraic identities---increasing its complexity while preserving its
value---and then \emph{unscramble} it back to canonical form step by
step, recording each step as a labeled training sample. This generates expert trajectories without requiring an existing solver.
One can then train a classifier to predict the 
correct reduction action (mathematical identity to apply) at each step from the set of all possible actions. In Ref.~\cite{Shih:2026lmy} we showed that this ``learning to unscramble" strategy was highly effective (nearly 100\% solve rate) at simplifying dilogarithm sums and spinor-helicity amplitudes. 

As noted in the introduction, however, IBP reduction presents a somewhat different challenge compared to symbolic simplification.
It is a \emph{weight reduction} problem, where the starting point (a single high-weight integral) is simpler than the endpoint (a linear combination of master integrals). 
Simply
reversing the scramble order as in Ref.~\cite{Shih:2026lmy} produces targets in an order unrelated to
integral weight, which provides a poor training signal for a weight reduction problem. We overcome these
challenges by exploiting the linearity of IBP identities to freely
reorder the unscramble sequence (see
Appendix~\ref{app:reordering_proof} for a formal proof), as we
describe below.

A key design choice that informs the training data generation is that
the MDP is trained to reduce integrals purely \emph{within} a given
sector, modulo subsectors. When reducing an integral in sector $S$, the
agent need only eliminate integrals belonging to $S$ that lie above the
corner integral; any subsector integrals (belonging to strict subsectors
$S' \subset S$) are treated as already reduced. In the hierarchical
reduction strategy (Section~\ref{sec:compeff}), subsector integrals are
handled independently by separate workers. This scoping is reflected in
the training data: scrambles are seeded within a single sector, and the
unscramble targets only integrals within that sector.

Concretely, for a given sector $S$, the starting expression $M$ is the
corner integral of $S$---the unique integral with all propagators of $S$
at unit power and no ISP insertions---multiplied by a random nonzero
coefficient in $\mathbb{Z}_p$. This expression is scrambled by applying
$N$ random IBP/LI identities: at each step, a random identity template
is selected and evaluated at a seed integral present in the expression,
the identity is solved for that integral, and the solution is
substituted into the expression. By the sector preservation property
(Section~\ref{sec:ibp_background}), identities seeded within sector $S$
produce integrals only in $S$ or its subsectors, so no explicit sector
filtering is needed during scrambling. After $N$ steps, the result is a
complex expression $\mathcal{E}_N$ involving integrals at various
weights within $S$ and its subsectors.

After scrambling, the expression can be written as
\begin{equation}
\mathcal{E}_N = M + \sum_{i=1}^{N} \alpha_i \, R_i, \quad \alpha_i \neq 0,
\label{eq:scramble_decomp}
\end{equation}
where $R_i$ is the IBP identity used in the $i$-th scramble step and
$\alpha_i$ is the nonzero coefficient determined by the substitution.
Since each $R_i \equiv 0$, we have $\mathcal{E}_N = M$ algebraically.
The simplest unscramble strategy, used in
Ref.~\cite{Shih:2026lmy}, reverses the scramble: at step $k$, use
$R_{N-k+1}$ to eliminate the integral it introduced. However, the
resulting targets follow the scramble order, which has no systematic
relationship to integral weight. Our key insight is that because IBP
identities are linear, the scramble operations commute:
$\mathcal{E}_N$ is the same regardless of the order in which the
$\alpha_i R_i$ are summed. We exploit this by unscrambling in
\emph{weight order}: at each step, we target the highest-weight integral
in sector $S$ that lies above the corner, and search through the
recorded operations to find one that eliminates it. (Subsector integrals
are left untouched.) The formal proof that a valid operation can always
be found is given in Appendix~\ref{app:reordering_proof}.

At each unscramble step, a training sample is recorded consisting of the
current state (expression, substitution history, target, sector mask),
the enumerated list of valid actions for the target, and the index of
the oracle (scramble-derived) action within that list. While raw IBP
identities respect sector preservation, the accumulated substitution
chain can introduce integrals from higher sectors, so explicit sector
filtering is applied during action enumeration to ensure only
sector-consistent actions are included.

Despite the weight-ordered unscramble, a gap remains between training
and inference: training episodes begin from complex multi-term scrambled
expressions, while inference begins from a single integral. The model is
never directly trained on inputs resembling the inference task. That the
trained model nevertheless achieves high reduction rates is the central
empirical result of this work.

\subsection{Action Space: Polyencoder Cross-Attention}
\label{sec:architecture}

The MDP framework requires a classifier that, given the current state, produces a probability distribution over valid actions. 

The most obvious approach would be a conventional classifier with a fixed output layer mapping the state to probabilities over a pre-enumerated action space. However, several features of the IBP reduction problem make this impractical. The full action space---$N_{\text{op}}$ templates times all possible seeds $\mathbf{s} \in \mathbb{Z}^{N_p+N_s}$---is combinatorially large and grows with the number of propagators and ISPs. The set of \emph{valid} actions is much smaller -- typically 10-100 actions for a given expression -- but changes dynamically at every step, depending on the current expression and accumulated substitutions. A standard output layer (e.g.\ an MLP) has a fixed number of neurons set at construction time---its dimensionality cannot depend on the input---so it cannot natively produce an output that matches the variably-sized valid action space for every input. Instead, one would have to use the full action space and apply a valid action mask that depends on the input. Then even if outputting to the full action space were computationally feasible, this would lead to a very sparse valid action space -- for any given expression, only a small fraction of the full action space is valid, so a fixed output layer would waste most of its capacity producing scores for irrelevant, permanently masked outputs.

Instead, we are led to a completely different ML architecture for the action scoring classifier model. Scoring a variable-size set of candidates given a fixed context is precisely the \emph{learning-to-rank} problem studied extensively in information retrieval~\cite{Nogueira:2019passagereranking, Humeau:2020polyencoders}. In that setting, a query (analogous to our expression state) is used to score a variable-size set of candidate documents (analogous to our valid actions). 
We follow the \emph{poly-encoder} design of \cite{Humeau:2020polyencoders}, which encodes the query and candidates with separate encoders, then uses \emph{cross-attention} between them, providing an architecture that is both expressive and computationally efficient.  In practice this means that rather than having a fixed output space like a conventional classifier, the poly-encoder takes both the expression (the query) and the action (the candidate) as input. Using cross-attention, it then scores each action given the expression.  All valid actions are scored in a single forward pass, and a softmax over the logits yields a normalized probability distribution over exactly the valid action set. This allows the same trained model to rank the valid actions for any input expressions, regardless of the size of the valid action space, with no wasted capacity on invalid, masked actions.

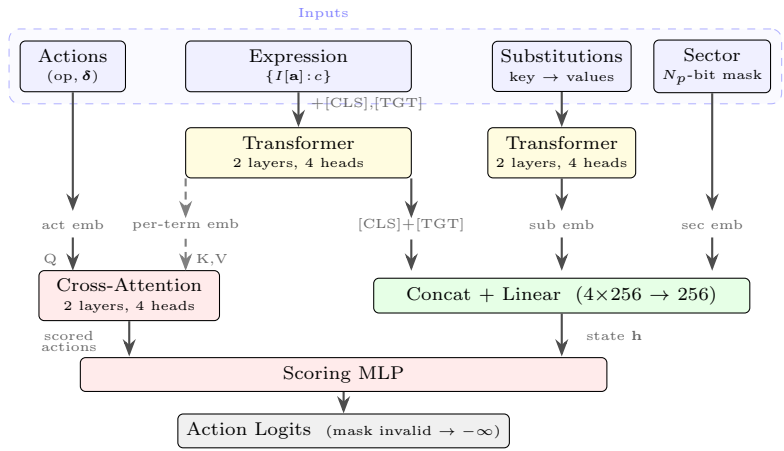
\begin{figure*}[tb]
\centering
\begin{tikzpicture}[
    >=Stealth,
    inputbox/.style={draw, rounded corners=2pt, fill=blue!6,
        minimum height=0.5cm, font=\scriptsize, align=center},
    transbox/.style={draw, rounded corners=2pt, fill=yellow!15,
        minimum height=0.45cm, font=\scriptsize, align=center},
    combbox/.style={draw, rounded corners=2pt, fill=green!10,
        minimum height=0.45cm, font=\scriptsize, align=center},
    scorebox/.style={draw, rounded corners=2pt, fill=red!8,
        minimum height=0.45cm, font=\scriptsize, align=center},
    outputbox/.style={draw, rounded corners=2pt, fill=gray!12,
        minimum height=0.45cm, font=\scriptsize, align=center},
    arr/.style={->, thick, color=black!70},
    darr/.style={->, thick, densely dashed, color=black!50},
    lbls/.style={font=\tiny, text=black!60},
]

\def\yIn{0}
\def\yTr{-1.15}
\def\yLbl{-2.1}
\def\yMid{-3.05}
\def\yMLP{-4.1}
\def\yOut{-4.85}

\node[inputbox, minimum width=1.4cm] (act_in) at (0, \yIn)
    {Actions\\[-1pt]{\tiny $(\text{op}, \boldsymbol{\delta})$}};
\node[inputbox, minimum width=3.0cm] (expr_in) at (3, \yIn)
    {Expression\\[-1pt]{\tiny $\{I[\mathbf{a}]\!:\!c\}$}};
\node[inputbox, minimum width=1.7cm] (sub_in) at (6.5, \yIn)
    {Substitutions\\[-1pt]{\tiny key $\to$ values}};
\node[inputbox, minimum width=1.3cm] (sec_in) at (8.5, \yIn)
    {Sector\\[-1pt]{\tiny $N_p$-bit mask}};

\node[transbox, minimum width=3.0cm] (expr_trans) at (3, \yTr)
    {Transformer\\[-1pt]{\tiny 2 layers, 4 heads}};
\node[transbox, minimum width=1.7cm] (sub_trans) at (6.5, \yTr)
    {Transformer\\[-1pt]{\tiny 2 layers, 4 heads}};

\draw[arr] (expr_in) -- (expr_trans);
\node[lbls, right] at ($(expr_in.south)+(0.05,-0.15)$)
    {{\tiny +[CLS],[TGT]}};
\draw[arr] (sub_in) -- (sub_trans);

\node[lbls] (act_lbl)     at (0,   \yLbl) {{\tiny act emb}};
\node[lbls] (perterm_lbl) at (1.5, \yLbl) {{\tiny per-term emb}};
\node[lbls] (cls_lbl)     at (4.5, \yLbl) {{\tiny [CLS]+[TGT]}};
\node[lbls] (sub_lbl)     at (6.5, \yLbl) {{\tiny sub emb}};
\node[lbls] (sec_lbl)     at (8.5, \yLbl) {{\tiny sec emb}};

\draw[arr] (act_in) -- (act_lbl);
\draw[darr] ($(expr_trans.south)+(-1.5,0)$) -- (1.5, \yLbl+0.08);
\draw[arr] ($(expr_trans.south)+(1.5,0)$) -- (4.5, \yLbl+0.08);
\draw[arr] (sub_trans) -- (sub_lbl);
\draw[arr] (sec_in) -- (sec_lbl);

\node[scorebox, minimum width=2.4cm] (crossattn) at (0.75, \yMid)
    {Cross-Attention\\[-1pt]{\tiny 2 layers, 4 heads}};
\node[combbox, minimum width=5.0cm] (concat) at (6.5, \yMid)
    {Concat + Linear~~(4$\times$256 $\to$ 256)};

\draw[arr] (act_lbl) -- (0, \yMid+0.33);
\node[lbls] at (-0.3, {(\yLbl+\yMid)/2}) {Q};
\draw[darr] (perterm_lbl) -- (1.5, \yMid+0.33);
\node[lbls] at (1.85, {(\yLbl+\yMid)/2}) {K,V};

\draw[arr] (cls_lbl) -- (4.5, \yMid+0.33);
\draw[arr] (sub_lbl) -- (6.5, \yMid+0.33);
\draw[arr] (sec_lbl) -- (8.5, \yMid+0.33);

\node[scorebox, minimum width=7.0cm] (mlp) at (3.6, \yMLP)
    {Scoring MLP};

\draw[arr] (crossattn) -- (0.75, \yMLP+0.24);
\node[lbls] at (-0.05, {(\yMid+\yMLP)/2}) {{\tiny scored}};
\node[lbls] at (-0.05, {(\yMid+\yMLP)/2-0.15}) {{\tiny actions}};

\draw[arr] (concat) -- (6.5, \yMLP+0.24);
\node[lbls] at (7.2, {(\yMid+\yMLP)/2}) {{\tiny state $\mathbf{h}$}};

\node[outputbox, minimum width=4.0cm] (output) at (3.6, \yOut)
    {Action Logits~~{\tiny (mask invalid $\to -\infty$)}};
\draw[arr] (mlp) -- (output);

\begin{pgfonlayer}{background}
    \node[draw=blue!30, dashed, rounded corners=4pt, fill=blue!2,
        fit=(act_in)(sec_in),
        inner sep=4pt, label={[font=\tiny, text=blue!50]above left:Inputs}] {};
\end{pgfonlayer}

\end{tikzpicture}
\caption{Architecture of the action classifier. All inputs are embedded to a common dimension ($d_{\text{emb}} = 256$); embedding layers are described in Appendix~\ref{app:embeddings}. Expression terms and substitution histories are each processed by independent Transformer encoders. The expression Transformer produces per-term embeddings (keys/values for cross-attention, dashed) and pooled \texttt{[CLS]}+\texttt{[TGT]} tokens. The global state vector $\mathbf{h}$ concatenates the \texttt{[CLS]}, \texttt{[TGT]}, substitution, and sector representations. Action scores are computed via cross-attention (actions as queries, expression terms as keys/values) followed by a scoring MLP that combines the attended action representations with $\mathbf{h}$.}
\label{fig:architecture}
\end{figure*}

Fig.~\ref{fig:architecture} shows a diagram of our action scoring architecture. The expression state---comprising the current expression, substitution history, target integral, and sector mask---is encoded by a Transformer without positional encoding, producing  a set of per-term expression embeddings $\{e_t\}_{t=1}^T$. (It also produces a global state vector $\mathbf{h}$; more on this below.) Each valid action $(a, \mathbf{s})$ is independently encoded into an embedding $\mathbf{a}_i$ by concatenating a learned template embedding with a seed encoding and projecting to dimension $d_{\text{emb}}$ (see Appendix~\ref{app:embeddings} for full details of all encoding modules). 

To score each action against the expression, we use cross-attention: each action embedding attends to the per-term expression embeddings to gather context about which expression terms are relevant to that particular action.\footnote{Note that in the attention mechanism, each \emph{candidate action} serves as the attention query---it ``looks up'' information in the expression context---which is the reverse of the information-retrieval convention where the context is called the ``query.''} More explicitly, the attention weights are
\begin{equation}
\alpha_{it} = \frac{\exp(q_i \cdot k_t / \sqrt{d})}{\sum_{t'=1}^{T} \exp(q_i \cdot k_{t'} / \sqrt{d})}
\label{eq:cross_attn}
\end{equation}
with $q_i = W_Q a_i$, $k_t = W_K e_t$, $v_t = W_V e_t$. These are used to form the attended action representations 
\begin{equation}
\label{eq:attendedaction}
\tilde{a}_i = \sum_{t} \alpha_{it}\, v_t
\end{equation} 
In our implementation, the cross-attention uses 4 heads and 2 layers (with layer normalization and feed-forward blocks between layers), allowing iterative refinement of the action representations conditioned on expression context. We emphasize that this architecture naturally respects the symmetries of the inputs. In particular, invariance under permutation of the terms in the input expression is guaranteed by the permutation symmetry of the attention mechanism. Since actions are scored individually, there is no spurious dependence of the scoring on the action indexing or ordering.

\subsection{Global state vector and final scoring}

Two special tokens are prepended to the expression sequence before it enters the Transformer. The first is a learnable \texttt{[CLS]} token (a trained parameter vector, independent of the input) that serves as a global summary of the expression, analogous to the \texttt{[CLS]} token in BERT~\cite{Devlin:2018mgb}. The second is a \texttt{[TARGET]} token that identifies the current reduction target (the highest-weight non-master integral, selected before scoring). The target integral is encoded using the same per-index embedding as expression integrals, but without a coefficient; note that the target also appears as one of the expression terms (with its coefficient), so it is effectively represented twice---once to mark it as the reduction target and once as part of the expression. Since the Transformer has no positional encoding, these special tokens are distinguished from expression terms not by their position in the sequence but by their embedding content. Both participate in self-attention on equal footing with the expression terms, allowing them to aggregate information from the full expression. Their output embeddings are projected and concatenated (along with the substitution and sector embeddings) to form the (permutation invariant) global state vector $\mathbf{h}$.

By combining the attended action representations $\tilde a_i$ from (\ref{eq:attendedaction}) with the global state vector in a final MLP layer, we obtain the final score for action $i$:
\begin{equation}
s_i = \mathrm{MLP}(\mathbf{h},\, \tilde{a}_i)
\end{equation}
As described above, all the valid actions are scored simultaneously this way, and then the scores are normalized with an overall softmax to form probabilities over the valid action space.

The classifier is then trained with cross-entropy loss on the expressions from each unscrambling step from Sec.~\ref{sec:trainingdata}, with the oracle actions as the target labels. This teaches the model to predict oracle actions that promote the reduction of high weight integrals to lower weight integrals.

\subsection{Reduction Strategy}

Attempting to fully reduce a high weight integral down to master integrals with a single continuous run of the MDP turns out to be unfeasible (at least with our current setup) -- there is too much of a gap between the training data (expressions with many terms containing a range of weights) and the target data (single high weight integrals). Instead, we require a more elaborate reduction strategy in order to complete the reduction down to masters.

The strategy has the following components:
\begin{itemize}
\item {\bf Hierarchical single-episode reduction}: we leverage the fact that the reduction can be made fully monotonic by the total ordering of loop integrals described in (\ref{eq:weight}). The reduction is defined relative to a set of \emph{goal states}---integrals that are considered fully reduced (e.g., corner integrals, master integrals, or any other chosen basis). It then suffices to reduce single integrals down to strictly lower-weight integrals in the same sector, plus goal-state integrals or subsector integrals of any weight. This we call one \emph{episode} of the reduction. Our reduction strategy targets the highest-weight non-goal-state integral in the expression, and after one episode, the next highest-weight integral is targeted, etc. As long as every episode successfully completes, the full reduction to goal states is guaranteed to succeed.

\item {\bf Beam search}: A simple greedy application of the MDP model (i.e.\ always applying the top ranked action to the expression) is insufficient for reducing high weight integrals even one episode. Instead we find it necessary to employ a beam search algorithm, where the top $K$ candidate states (according to a sorting criterion) are maintained over each MDP step of a single episode. At each step:

\begin{enumerate}
\item For each beam state, identify the highest-weight non-master target integral and enumerate valid actions.
\item Batch all (state, target, actions) tuples and run the classifier in a single forward pass.
\item For each state, extract the top-$K$ actions by model score, apply them in parallel to produce candidate next states.
\item Select the best $K$ states for the next beam according to the sorting criterion.
\end{enumerate}

We use a \emph{mixed} beam sort strategy that maintains two parallel beams of width $K$ each: one sorted by maximum integral weight (favoring aggressive weight reduction) and one sorted by total weight (favoring overall simplification). After deduplication, the total number of active states is between $K$ and $2K$. We take $K = 20$, so up to 40 states are maintained per step.

\end{itemize}

\subsection{Further Optimizations for Computational Efficiency}
\label{sec:compeff}

A key realization that greatly speeds up the hierarchical reduction is that since episodes operate on the level of single integrals, they can all be carried out independently and in parallel. We use a local HTCondor cluster (${\sim}$330 available CPU slots) to submit each episode as a single-CPU job. In more detail, the setup is:

\begin{enumerate}
\item An \emph{orchestrator} maintains the global expression and a cache of solved integrals.  \item For each non-master integral that appears in the expression, the orchestrator submits a \emph{one-episode worker} job that reduces the integral's weight by one level.
\item {\bf Worker isolation}: Crucially, each one-step worker begins with an \emph{empty} substitution history. The worker does not inherit the accumulated substitutions from the orchestrator or from prior workers. It builds up its own substitution history from scratch during the beam search, which it needs for discovering indirect actions within that episode. Once the worker completes and returns the solved integral's expression, the substitution history is discarded. This scoping is what gives the approach its bounded-memory property: no matter how many integrals have been reduced globally, each worker's memory footprint depends only on the beam width and the local reduction complexity, not on the total problem size.
\item When a worker completes, its result (the solved integral expressed in terms of lower-weight and/or lower-sector integrals) is cached.
\item {\bf Memoization}: When the same integral appears in a subsequent episode, the cached result is reused without recomputation. This {\it memoization} is critical for efficiency: in our benchmarks, cache hit rates range from 60\% to 75\%, meaning the majority of substitutions reuse previously computed results.
\item {\bf Asynchronous execution}: The orchestrator polls for completed jobs every 5 seconds, submitting new jobs as non-master integrals are discovered. There are no synchronization barriers between dependency levels; jobs at different levels of the reduction hierarchy run concurrently.
\item The process repeats until all non-master integrals are eliminated.
\end{enumerate}

To limit peak memory on each worker, model inference within each beam search step is sub-batched: the (state, target, actions) tuples are processed in groups of 50 through the classifier rather than all at once, preventing large intermediate tensors from accumulating. All inference is performed on CPU. Action enumeration and application are parallelized across 16 CPU cores per worker.

\section{Two-Loop Triangle-Box}
\label{sec:triangle_box}

As in Ref.~\cite{vonHippel:2025okr}, we choose the two-loop triangle-box integral family with massless internal propagators and massive external legs ($p_i^2 = m_i^2 \neq 0$) to serve as our concrete benchmark with which we demonstrate the framework above. It gives a moderately-sized system of master integrals and IBP+LI identities which is enough to furnish a non-trivial test of the framework, while still being computationally feasible for an initial proof-of-concept study.

\subsection{Topology and Kinematics}
\label{sec:topology}

The family is defined by six propagators and one ISP:
\begin{align}
D_1 &= k_1^2, &D_2 &= k_2^2, \notag \\
D_3 &= (k_1{+}k_2)^2, &D_4 &= (k_1{+}p_1)^2, \notag \\
D_5 &= (k_2{+}p_3)^2, &D_6 &= (k_2{-}p_1)^2, \notag \\
D_7 &= (k_1{+}p_3)^2 \;\; \text{(ISP)},
\label{eq:propagators}
\end{align}
where $k_1, k_2$ are loop momenta and $p_1, p_2, p_3$ are external momenta with $p_i^2 = m_i^2$. A general integral is
\begin{equation}
I[a_0, a_1, a_2, a_3, a_4, a_5, a_6] = \int \frac{d^d k_1 \, d^d k_2}{D_1^{a_0} D_2^{a_1} \cdots D_7^{a_6}},
\label{eq:integral}
\end{equation}
where $a_0, \ldots, a_5$ are propagator powers and $a_6 \leq 0$ encodes ISP insertions. The family has $2^6 - 1 = 63$ non-trivial sectors; the top sector (all six propagators present) has ID $63 = (111111)_2$. There are 16 master integrals distributed across 13 sectors~\cite{vonHippel:2025okr}:
\begin{align}
&I[0,0,1,1,1,0,0], &\quad& I[0,1,1,1,0,0,0], \nonumber \\
&I[0,1,1,1,1,0,0], && I[\text{-}1,1,1,1,1,0,0], \nonumber \\
&I[1,0,0,1,1,1,0], && I[1,0,1,0,0,1,0], \nonumber \\
&I[1,0,1,0,1,0,0], && I[1,0,1,0,1,1,0], \nonumber \\
&I[1,\text{-}1,1,0,1,1,0], && I[1,0,1,1,1,0,0], \nonumber \\
&I[1,\text{-}1,1,1,1,0,0], && I[1,0,1,1,1,1,0], \nonumber \\
&I[1,1,0,1,0,1,0], && I[1,1,0,1,1,0,0], \nonumber \\
&I[1,1,0,1,1,1,0], && I[1,1,1,1,1,0,0].
\label{eq:masters}
\end{align}
Note that not all masters are corner integrals: three have a negative index ($a_i = -1$).

The 8 IBP operators (from $\partial/\partial k_i \cdot v$ with $v \in \{k_1, k_2, p_1, p_2\}$ and $i \in \{1,2\}$) plus 1 LI identity give 9 equation templates; their explicit forms are listed in Appendix~\ref{app:identities}. Sector preservation is verified explicitly for all 9 operators there.

Following the finite-field approach now standard in IBP reduction~\cite{vonManteuffel:2014ixa,Peraro:2019svx,Klappert:2019emp}, we work with numerical kinematics and perform all arithmetic in $\mathbb{Z}_p$. We choose $p = 1009$, $d = 41$, $m_1 = p_1^2 = 1$, $m_2 = p_2^2 = 31$, $m_3 = p_3^2 = 47$. The key advantage of finite-field arithmetic is that all coefficients are single machine-precision integers, avoiding the intermediate expression swell that plagues exact rational arithmetic. Since the IBP+LI identity coefficients are rational functions of $d$ and the masses, a reduction path found for one numerical specialization generically holds for all values---the only exceptions are measure-zero loci where a pivot coefficient happens to vanish, which are avoided by choosing generic numerical values.

\subsection{Architecture and Training}
\label{sec:mdp}

For the triangle-box topology, the MDP of Section~\ref{sec:framework} has state coefficients in $\mathbb{Z}_p$, a 6-bit sector mask, and 9 IBP/LI templates (Section~\ref{sec:topology}) giving $a \in \{1, \ldots, 9\}$, with seeds $\mathbf{s} \in \mathbb{Z}^7$. Typically 10--100 valid actions exist per state.

Following the general poly-encoder design described in Section~\ref{sec:architecture}, the state is encoded into four input groups (see Appendix~\ref{app:embeddings} for details): (i)~expression terms, processed by a 2-layer, 4-head Transformer encoder with prepended \texttt{[CLS]} and \texttt{[TARGET]} tokens; (ii)~the substitution history (up to 50 entries), encoded via attention-pooling over replacement terms and processed by a separate 2-layer Transformer; (iii)~the 6-bit sector mask, encoded via per-bit embeddings; and (iv)~actions $(a, \mathbf{s})$, encoded by embedding the template index and seed. All inputs are embedded to a common dimension $d_{\text{emb}} = 256$. The \texttt{[CLS]} output, \texttt{[TARGET]} output, substitution embedding, and sector embedding are concatenated ($4 \times 256 = 1024$ dimensions) and projected to the 256-dimensional state vector $\mathbf{h}$. Action embeddings attend to per-term expression embeddings via 2-layer multi-head cross-attention (4 heads), and the resulting attended representations are combined with $\mathbf{h}$ in a scoring MLP to produce per-action logits. Invalid actions are masked to $-\infty$. The full architecture is shown in Fig.~\ref{fig:architecture}.

Training data is generated by the scramble-and-unscramble procedure of Section~\ref{sec:framework}, run across all 63 non-trivial sectors with $n \in [5, 20]$ random IBP operations per scramble. We generate ${\sim}8 \times 10^4$ trajectories, yielding ${\sim}1.06 \times 10^6$ training samples (with a held-out validation set of ${\sim}118$K samples).

As illustrative examples, a typical training sample from the early steps of a trajectory in sector $(001000)_2$ might be:
\begin{align}
\text{expr} &= 719 \cdot I[0,0,1,0,0,0,0] + 1 \cdot I[0,0,2,-1,0,0,0] \notag \\
&\quad + 1008 \cdot I[0,0,2,0,0,-1,0], \notag \\
\text{target} &= I[0,0,2,0,0,-1,0], \quad w = (2,1), \notag \\
\text{oracle} &= (a{=}3,\; \mathbf{s} = (0,0,1,0,0,0,0)),
\label{eq:train_ex1}
\end{align}
where the oracle action applies the template $\partial/\partial k_1^\mu \cdot (k_1{+}k_2)^\mu$ to the seed $I[0,0,1,0,0,0,0]$.
A sample from a higher-weight trajectory in sector $(110011)_2$ might be:
\begin{align}
\text{expr} &= 205 \cdot I[1,1,0,0,1,1,0] + 991 \cdot I[1,1,0,0,2,1,0] \notag \\
&\quad + 1 \cdot I[2,1,-1,0,1,2,0] + 65 \cdot I[2,1,0,-1,2,1,0] \notag \\
&\quad + 944 \cdot I[2,1,0,0,2,1,0], \notag \\
\text{target} &= I[2,1,0,-1,2,1,0], \quad w = (6,1), \notag \\
\text{oracle} &= (a{=}3,\; \mathbf{s} = (1,1,0,0,2,1,0)),
\label{eq:train_ex2}
\end{align}
where the oracle action applies the same template to the seed $I[1,1,0,0,2,1,0]$. In each case, the classifier must select the oracle action from among typically 50--100 valid alternatives. Recall that all coefficients are in $\mathbb{Z}_p$ with $p = 1009$.

\begin{figure}[tb]
\centering
\includegraphics[width=\columnwidth]{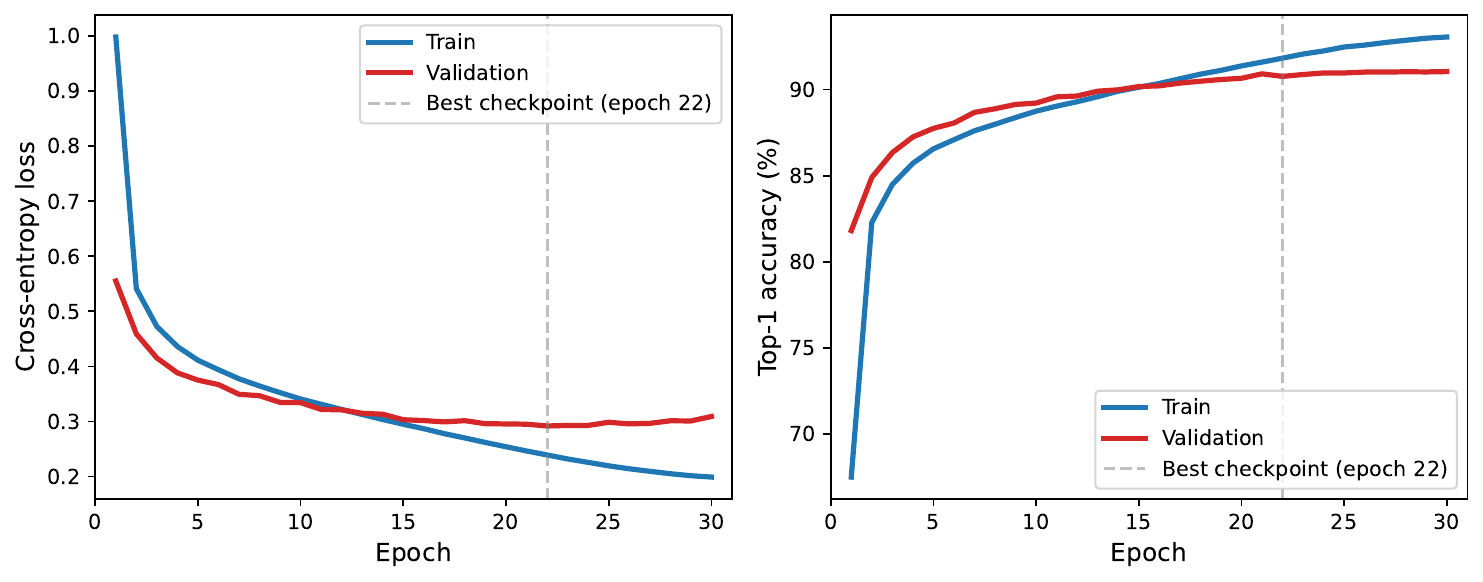}
\caption{Training and validation cross-entropy loss (left) and top-1 accuracy (right) as a function of epoch. The dashed line marks epoch 22, where the best validation loss is achieved.}
\label{fig:training_curve}
\end{figure}

The model has approximately 7.7M trainable parameters and is trained for 30 epochs with cross-entropy loss, AdamW optimizer (learning rate $4 \times 10^{-4}$, weight decay $10^{-5}$), cosine annealing, gradient clipping at 1.0, and batch size 256. The best checkpoint (epoch 22) achieves 90.8\% top-1 accuracy and ${\sim}$99\% top-5 accuracy on the validation set, indicating that the model learns to identify the oracle action with high reliability. The training and validation loss and accuracy curves are shown in Fig.~\ref{fig:training_curve}.

\subsection{Results}
\label{sec:results}

We benchmark \textsc{Sailir} against \prog{Kira} 3.1 (with default settings), one of the leading implementations of the Laporta algorithm, which combines finite-field reconstruction with algebraic simplification to achieve state-of-the-art performance.  All \textsc{Sailir} benchmarks use a single trained model checkpoint (epoch 22) with beam width $K = 20$, prime $p = 1009$, and mixed beam sort.

We compare on 16 integrals from the two-loop triangle-box family, spanning total positive index weight $r \in \{10, \ldots, 13\}$ and total numerator power $s \in \{4, \ldots, 7\}$ (Section~\ref{sec:topology}). All benchmarked integrals are in the top sector; for each $(r, s)$, the propagator indices were chosen randomly subject to the constraint $\sum_i a_i = r$ and $a_6 = -s$. The 16 benchmark integrals are listed in Table~\ref{tab:integrals}. 

\begin{table}[tb]
\centering
\small
\begin{tabular}{|r|r|l||r|r|l|}
\hline
$r$ & $s$ & Integral & $r$ & $s$ & Integral \\
\hline
10 & 4 & $I[2,1,2,1,2,2,-4]$ & 10 & 6 & $I[1,1,3,2,2,1,-6]$ \\
11 & 4 & $I[2,2,2,1,1,3,-4]$ & 11 & 6 & $I[1,4,2,1,2,1,-6]$ \\
12 & 4 & $I[2,3,1,3,1,2,-4]$ & 12 & 6 & $I[3,2,3,2,1,1,-6]$ \\
13 & 4 & $I[2,3,3,3,1,1,-4]$ & 13 & 6 & $I[3,2,1,3,2,2,-6]$ \\
\hline
10 & 5 & $I[1,1,2,2,1,3,-5]$ & 10 & 7 & $I[2,3,1,1,2,1,-7]$ \\
11 & 5 & $I[1,1,2,3,2,2,-5]$ & 11 & 7 & $I[2,1,1,2,3,2,-7]$ \\
12 & 5 & $I[1,2,2,2,1,4,-5]$ & 12 & 7 & $I[3,1,1,1,1,5,-7]$ \\
13 & 5 & $I[2,2,3,3,2,1,-5]$ & 13 & 7 & $I[2,2,3,3,1,2,-7]$ \\
\hline
\end{tabular}
\caption{The 16 benchmark integrals, all in the top sector $(111111)_2$. For each $(r, s)$, the propagator indices were chosen randomly.}
\label{tab:integrals}
\end{table}

\begin{figure*}[tb]
\centering
\includegraphics[width=\textwidth]{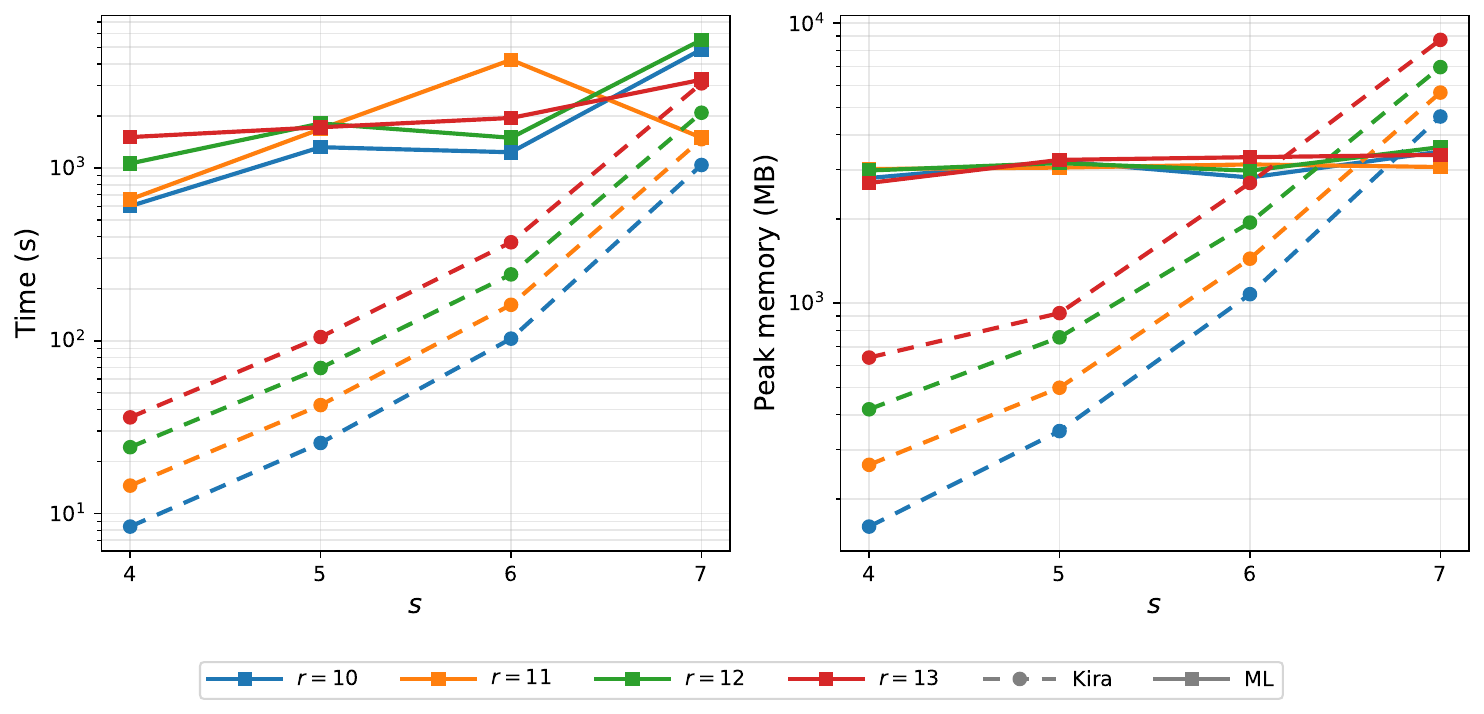}
\caption{\prog{Kira} vs.\ \textsc{Sailir} on the two-loop triangle-box topology, for $r \in \{10, \ldots, 13\}$ and $s \in \{4, \ldots, 7\}$. Dashed lines with circles: \prog{Kira}. Solid lines with squares: \textsc{Sailir} (ideal parallel time, per-worker memory). Colors indicate different values of $r$. \textbf{Left}: Time vs.\ $s$. \prog{Kira} time grows rapidly with $s$; \textsc{Sailir} time grows more slowly, and the curves converge at high $s$. \textbf{Right}: Peak memory vs.\ $s$. \prog{Kira} memory grows by nearly two orders of magnitude; \textsc{Sailir} per-worker memory remains approximately flat at ${\sim}$3~GB. The memory crossover occurs between $s = 6$ and $s = 7$.}
\label{fig:benchmark}
\end{figure*}

For the reduction, we set the goal states to be the 16 master integrals in Eq.~(\ref{eq:masters}). The complete per-integral results are given in Table~\ref{tab:benchmark} in Appendix~\ref{app:benchmark_table}. All 16 integrals are successfully reduced to master integrals by both methods.

The fact that the model was able to reduce to 16 master integrals, despite not being trained on this objective (recall from Section~\ref{sec:trainingdata} that it was trained on scrambles of the 63 \emph{corner} integrals) is highly nontrivial. It indicates that the model was able to generalize beyond its training data  and learn the underlying structure of IBP reduction rather than memorizing specific reduction paths.

Shown in Fig.~\ref{fig:benchmark} are the memory and timing comparisons between \textsc{Sailir} and \prog{Kira}. 
The most striking result is the contrasting memory scaling behavior, shown in the right panel of Fig.~\ref{fig:benchmark}. \prog{Kira}'s peak memory grows rapidly with both $r$ and $s$, increasing by nearly two orders of magnitude from 159~MB ($r{=}10$, $s{=}4$) to 8.7~GB ($r{=}13$, $s{=}7$). This reflects the fundamental scaling of the Laporta approach: more complex integrals require larger systems of equations, and all equations must be stored simultaneously.

In contrast, \textsc{Sailir}'s per-worker memory remains approximately flat at ${\sim}$2.7--3.6~GB across all 16 integrals, independent of complexity. This is because each worker reduces a single integral by one weight level, and the worker's memory consumption is dominated by the model size and beam search data structures, not by the integral's complexity. 

The memory crossover occurs between $s = 6$ and $s = 7$: for $s \leq 6$, \prog{Kira} uses less memory; for $s = 7$, \textsc{Sailir} uses only 0.4--0.7$\times$ as much memory as \prog{Kira}. Extrapolating to higher $s$ values, the memory advantage of \textsc{Sailir} is expected to grow further, since \textsc{Sailir} memory remains flat while \prog{Kira} memory continues to increase.

\textsc{Sailir} is slower than \prog{Kira} for all tested integrals, with the time ratio (\textsc{Sailir}/\prog{Kira}) decreasing from ${\sim}$72$\times$ at the simplest integral ($r{=}10$, $s{=}4$) to ${\sim}$1.0$\times$ at the most complex.\footnote{The time reported here for \textsc{Sailir} is the {\it ideal parallel time}:  the length of the longest chain of sequential dependencies, where each link is weighted by the actual worker computation time. This represents the minimum wall-clock time achievable with unlimited parallelism and zero scheduling overhead. In practice, the actual wall-clock time is typically 1.5--2$\times$ the ideal parallel time, with the overhead coming from Condor job scheduling latency and the orchestrator's 5-second polling interval. With a dedicated parallel execution framework (rather than batch scheduling), this overhead could be largely eliminated.} This convergence is visible in the left panel of Fig.~\ref{fig:benchmark}: \prog{Kira}'s time grows roughly exponentially with complexity, while the \textsc{Sailir} ideal parallel time grows more slowly. The two curves approach each other at $s = 7$, with the closest time ratios being 1.0$\times$ ($r{=}11$, $s{=}7$) and 1.0$\times$ ($r{=}13$, $s{=}7$).

The \textsc{Sailir} time scaling is driven by the number of unique integrals encountered during reduction (reflected in the ``Jobs'' column of Table~\ref{tab:benchmark}), which grows with $r$ and $s$ but more slowly than the system size in the Laporta approach. The cache hit rate ranges from 60\% to 75\%, indicating substantial reuse of computed results across different reduction paths.

\section{Discussion and Conclusions}
\label{sec:discussion}

We have presented \textsc{Sailir}, a self-supervised AI approach to IBP reduction that achieves bounded per-worker memory consumption, in contrast to the rapidly growing memory requirements of traditional Laporta-based methods. The key ingredients are: (1)~a self-supervised training procedure based on the scramble-and-unscramble framework of Ref.~\cite{Shih:2026lmy}, which generates expert oracle trajectories without requiring an existing IBP solver; (2)~a poly-encoder action classifier based on cross-attention, which efficiently scores variable-sized sets of valid actions---essential because the full action space is combinatorially large and a fixed-output architecture would be infeasible; and (3)~a hierarchical, single-episode reduction strategy with asynchronous parallel execution and memoization, where each worker reduces one integral by one weight level with bounded memory, and results are cached and reused across the reduction tree.

On the two-loop triangle-box topology, \textsc{Sailir} successfully reduces all 16 benchmark integrals to master integrals, with per-worker memory remaining flat at ${\sim}$3~GB regardless of integral complexity. For the most complex integrals ($s = 7$), this represents a 2--3$\times$ memory savings compared to \prog{Kira}, and the advantage is expected to grow for more complex integrals and topologies. The timing overhead is significant for simple integrals (where \prog{Kira} takes only seconds) but converges toward parity for complex ones (where \prog{Kira} takes ${\sim}$50 minutes).

Several directions for improvement and extension are worth noting:

\textbf{Scaling to harder topologies.} The two-loop triangle-box is a relatively simple topology with 6 propagators, 16 masters, and modest index values. Current frontier applications involve topologies with more propagators, hundreds of masters, and higher index values. \textsc{Sailir} should in principle scale better than Laporta to such cases, since the per-worker memory remains bounded. The model will need to be retrained for each new topology, but this is a one-time cost per topology and then all integrals from that family can be reduced using the same model. Importantly, the self-supervised training procedure requires no prior reduction data---training samples are generated entirely from scrambling corner integrals of the new topology---so \textsc{Sailir} can be applied to any topology for which the IBP+LI identities can be written down, including ones that have never been reduced before.

\textbf{Better models.} The current model is a relatively small transformer (${\sim}$6--8M parameters) trained for 30 epochs on ${\sim}$50--100K samples. Larger models, more training data (from more scrambles and higher complexity), and techniques such as curriculum learning could improve the model's accuracy and reduce the number of beam search steps needed.

\textbf{Reducing time overhead.} The bulk of the \textsc{Sailir} time is spent on action enumeration (checking which IBP+LI identities are applicable) and model inference. Faster action enumeration (via better caching or precomputation), GPU-based model inference, and a dedicated parallel execution framework (rather than Condor batch scheduling) could substantially reduce the time gap with \prog{Kira}.

\textbf{Hybrid approaches.} \textsc{Sailir} and Laporta have complementary strengths: Laporta is fast for simple integrals, while \textsc{Sailir} has bounded memory for complex ones. A hybrid strategy that uses Laporta for simple sectors and \textsc{Sailir} for the most memory-intensive reductions could combine the best of both worlds.

The broader significance of this work is demonstrating that sequential, AI-guided decision-making can serve as a viable alternative to global linear algebra for IBP reduction, with qualitatively different resource scaling. As perturbative calculations continue to push toward higher complexity, the bounded-memory property of \textsc{Sailir} may become increasingly valuable.

\section*{Acknowledgements}
I am grateful to Tommaso Armadillo, Fabrizio Caola, Aurelien Dersy, Federica Devoto, Laura Reina for helpful discussions, and to Tommaso Armadillo, Federica Devoto and Mao Zeng for providing feedback on the draft. This work was done in full collaboration with Claude Code Opus 4.5-4.6. Claude did all of the hands on work under my supervision -- writing the code; training, evaluating and optimizing the ML algorithms; analyzing the results; making plots and tables; and contributing significantly to the writing of the paper. This research was supported by DOE grant DOE-SC0010008. This work was performed in part at the Aspen Center for Physics, which is supported by National Science Foundation grant PHY-2210452.

\section*{Code}
Our code is available at \url{https://github.com/davidshih17/RL_IBPreduction_claude}.

\appendix

\section{Proof of the Validity of the Highest-Weight Unscrambling Strategy}
\label{app:reordering_proof}

Here we give the proof of the claim in Sec.~\ref{sec:trainingdata} that for any scrambled expression, targeting the highest weight integral and picking an action from the list of IBP operations used in the scrambling will always succeed in unscrambling the expression.

In more detail, the procedure is: at each step, identify the highest-weight non-master integral in the expression, find one of the recorded IBP+LI identities that contains it (after applying all accumulated substitutions), use that identity to solve for the integral, and substitute the solution into the expression---thereby eliminating it. This is analogous to using a row of a matrix to eliminate a variable in Gaussian elimination.

Formally, after $k$ such steps using a subset $U_k$ of the recorded operations, the expression maintains the invariant
\begin{equation}
\mathcal{E}_k = M + \sum_{j \notin U_k} \alpha_j \, \tilde{R}_j^{(k)},
\label{eq:invariant}
\end{equation}
where $\tilde{R}_j^{(k)}$ denotes the equation $R_j$ after applying all $k$ accumulated substitutions. Each substitution eliminates one target integral from the expression: when equation $R_i$ is used to solve for its target, the coefficient of $R_i$ in the sum cancels exactly, so used equations drop out. Crucially, the unused equations retain their \emph{original} coefficients $\alpha_j \neq 0$: substitutions transform $R_j$ into $\tilde{R}_j^{(k)}$ but do not alter $\alpha_j$. Since the sum is non-zero, at least one unused equation $\tilde{R}_j^{(k)}$ must contain a non-zero term, ensuring a valid pivot always exists. After all $N$ equations are used, the sum is empty and $\mathcal{E}_N = M$. This is algebraically equivalent to Gaussian elimination with flexible row selection.

\section{Explicit IBP and LI Identities}
\label{app:identities}

We list the 8 IBP identities and 1 Lorentz invariance (LI) identity for the 2-loop triangle box topology used in this work. Each identity is a linear relation among integrals $I[\mathbf{a}]$ with shifted indices, with coefficients that are polynomials in the seed indices $a_0, \ldots, a_6$, the spacetime dimension $d$, and the external masses $m_2 = p_2^2$, $m_3 = p_3^2$. The mass $m_1 = p_1^2 = 1$ has been substituted throughout.

\textbf{Notation.} We use raising and lowering operators: $j^+$ shifts $a_j \to a_j + 1$ and $j^-$ shifts $a_j \to a_j - 1$, both acting on $I[\mathbf{a}]$. Products of operators compose: $0^- 2^+$ denotes $I[a_0{-}1, a_1, a_2{+}1, a_3, a_4, a_5, a_6]$; by convention, the lowering operator is written first. The identity operator (no shift) is~$\mathbbm{1}$.

At runtime, all coefficients are evaluated with numerical kinematics $d = 41$, $m_2 = 31$, $m_3 = 47$, and reduced modulo $p = 1009$.

\allowdisplaybreaks

\textbf{Op 0} ($\partial/\partial k_1 \cdot k_1$, 7 terms):
\begin{align}
0 &= (d{-}2a_0{-}a_2{-}a_3{-}a_6)\,\mathbbm{1}  - a_2\,0^{\!-}2^{\!+} + a_2\,1^{\!-}2^{\!+} \notag \\
&\quad + m_3 a_6\,6^{\!+} - a_3\,0^{\!-}3^{\!+} - a_6\,0^{\!-}6^{\!+} + a_3\,3^{\!+}
\end{align}

\textbf{Op 1} ($\partial/\partial k_2 \cdot k_2$, 7 terms):
\begin{align}
0 &= (d{-}2a_1{-}a_2{-}a_4{-}a_5)\,\mathbbm{1} + a_2\,0^{\!-}2^{\!+} - a_2\,1^{\!-}2^{\!+} \notag \\
&\quad + a_5\,5^{\!+} - a_4\,1^{\!-}4^{\!+} - a_5\,1^{\!-}5^{\!+} + m_3 a_4\,4^{\!+}
\end{align}

\textbf{Op 2} ($\partial/\partial k_1 \cdot p_1$, 12 terms):
\begin{align}
0 &= (a_0{-}a_3)\,\mathbbm{1} - a_0\,3^{\!-}0^{\!+} + a_0\,0^{\!+} + a_2\,0^{\!-}2^{\!+} \notag \\
&\quad - a_2\,1^{\!-}2^{\!+} - a_2\,3^{\!-}2^{\!+} + a_2\,5^{\!-}2^{\!+} + a_3\,0^{\!-}3^{\!+} \notag \\
&\quad + (m_2{-}m_3) a_6\,6^{\!+} - a_3\,3^{\!+} + a_6\,0^{\!-}6^{\!+} - a_6\,3^{\!-}6^{\!+}
\end{align}

\textbf{Op 3} ($\partial/\partial k_2 \cdot p_1$, 12 terms):
\begin{align}
0 &= (a_5{-}a_1)\,\mathbbm{1} + a_1\,5^{\!-}1^{\!+} - a_1\,1^{\!+} + a_2\,0^{\!-}2^{\!+} \notag \\
&\quad - a_2\,1^{\!-}2^{\!+} - a_2\,3^{\!-}2^{\!+} + a_2\,5^{\!-}2^{\!+} - a_4\,1^{\!-}4^{\!+} \notag \\
&\quad + a_4\,5^{\!-}4^{\!+} + (m_2{-}m_3{-}2) a_4\,4^{\!+} - a_5\,1^{\!-}5^{\!+} + a_5\,5^{\!+}
\end{align}

\textbf{Op 4} ($\partial/\partial k_2 \cdot k_1$, 14 terms):
\begin{align}
0 &= (a_1{-}a_2)\,\mathbbm{1} + a_1\,0^{\!-}1^{\!+} - a_1\,2^{\!-}1^{\!+} - a_2\,0^{\!-}2^{\!+} \notag \\
&\quad + a_2\,1^{\!-}2^{\!+} - a_5\,5^{\!+} + 2a_4\,0^{\!-}4^{\!+} + a_4\,1^{\!-}4^{\!+} \notag \\
&\quad - a_4\,2^{\!-}4^{\!+} - a_4\,6^{\!-}4^{\!+} + m_3 a_4\,4^{\!+} + a_5\,1^{\!-}5^{\!+} \notag \\
&\quad - a_5\,2^{\!-}5^{\!+} + a_5\,3^{\!-}5^{\!+}
\end{align}

\textbf{Op 5} ($\partial/\partial k_1 \cdot k_2$, 14 terms):
\begin{align}
0 &= (a_0{-}a_2)\,\mathbbm{1} + a_0\,1^{\!-}0^{\!+} - a_0\,2^{\!-}0^{\!+} + a_2\,0^{\!-}2^{\!+} \notag \\
&\quad - a_2\,1^{\!-}2^{\!+} + m_3 a_6\,6^{\!+} + a_3\,0^{\!-}3^{\!+} - a_3\,2^{\!-}3^{\!+} \notag \\
&\quad + a_3\,5^{\!-}3^{\!+} - a_3\,3^{\!+} + a_6\,0^{\!-}6^{\!+} + 2a_6\,1^{\!-}6^{\!+} \notag \\
&\quad - a_6\,2^{\!-}6^{\!+} - a_6\,4^{\!-}6^{\!+}
\end{align}

\textbf{Op 6} ($\partial/\partial k_1 \cdot p_2$, 14 terms):
\begin{align}
0 &= (a_3{-}a_6)\,\mathbbm{1} + a_0\,3^{\!-}0^{\!+} - a_0\,6^{\!-}0^{\!+} + (m_3{-}1) a_0\,0^{\!+} \notag \\
&\quad + 2a_2\,1^{\!-}2^{\!+} + a_2\,3^{\!-}2^{\!+} - a_2\,4^{\!-}2^{\!+} - a_2\,5^{\!-}2^{\!+} \notag \\
&\quad - a_2\,6^{\!-}2^{\!+} + 2m_3 a_2\,2^{\!+} - a_3\,6^{\!-}3^{\!+} + m_2 a_3\,3^{\!+} \notag \\
&\quad + a_6\,3^{\!-}6^{\!+} - m_2 a_6\,6^{\!+}
\end{align}

\textbf{Op 7} ($\partial/\partial k_2 \cdot p_2$, 16 terms):
\begin{align}
0 &= (2a_1{-}a_4{-}a_5)\,\mathbbm{1} - a_1\,4^{\!-}1^{\!+} - a_1\,5^{\!-}1^{\!+} \notag \\
&\quad + (m_3{+}1) a_1\,1^{\!+} + 2a_2\,1^{\!-}2^{\!+} + a_2\,3^{\!-}2^{\!+} - a_2\,4^{\!-}2^{\!+} \notag \\
&\quad - a_2\,5^{\!-}2^{\!+} - a_2\,6^{\!-}2^{\!+} + 2m_3 a_2\,2^{\!+} + 2a_4\,1^{\!-}4^{\!+} \notag \\
&\quad + (2m_3{-}m_2) a_5\,5^{\!+} - a_4\,5^{\!-}4^{\!+} + (2{-}m_2) a_4\,4^{\!+} \notag \\
&\quad + 2a_5\,1^{\!-}5^{\!+} - a_5\,4^{\!-}5^{\!+}
\end{align}

\textbf{Op 8} (Lorentz invariance, 13 terms):
\begin{align}
0 &= (m_3{-}m_2{-}1) a_3\,0^{\!-}3^{\!+} + (1{-}m_2{-}m_3) a_3\,3^{\!+} \notag \\
&\quad + 2a_3\,6^{\!-}3^{\!+} + (m_2 m_3{+}m_3{-}m_3^2) a_6\,6^{\!+} \notag \\
&\quad + (m_3{+}m_2{-}1) a_6\,0^{\!-}6^{\!+} - 2m_3 a_6\,3^{\!-}6^{\!+} \notag \\
&\quad + 2m_3 a_4\,5^{\!-}4^{\!+} + (m_2{-}3m_3{-}1) a_4\,1^{\!-}4^{\!+} \notag \\
&\quad + (m_2 m_3{-}3m_3{-}m_3^2) a_4\,4^{\!+} - 2a_5\,4^{\!-}5^{\!+} \notag \\
&\quad + (m_3{-}m_2{+}3) a_5\,1^{\!-}5^{\!+} + (3m_3{-}m_2{+}1) a_5\,5^{\!+} \notag \\
&\quad + [(m_2{-}m_3{-}1)(a_3{+}a_5) + (m_3{-}m_2{+}1)(a_4{+}a_6)]\,\mathbbm{1}
\end{align}

{\bf Sector preservation}: Inspection of all 9 identities reveals a universal pattern: every term involving a raising operator $j^+$ (for $j = 0, \ldots, 6$) has a coefficient proportional to $a_j$. This is a direct consequence of the chain rule---differentiating $D_j^{-a_j}$ in the integrand produces $-a_j \cdot D_j^{-(a_j+1)}$---and holds for both IBP and LI identities. When $a_j = 0$ (propagator $D_j$ is absent from the seed), all $j^+$ terms vanish, so no IBP or LI equation can introduce a propagator that was not already present. This guarantees that IBP equations evaluated at a seed in sector $S$ produce integrals only in $S$ or its subsectors, as stated in Section~\ref{sec:ibp_background}.

\section{Embedding Layer Details}
\label{app:embeddings}

We describe the encoding modules summarized in Section~\ref{sec:mdp}. All modules output vectors of dimension $d_{\text{emb}} = 256$.

\textbf{IntegralEncoder.}
Each integral index tuple $(a_1, \ldots, a_{N_p + N_s})$ is encoded by combining per-position embeddings with explicit weight features. Each index position $i$ has its own learned embedding table mapping integer values to vectors of dimension $d_{\text{emb}}/2$. In parallel, the weight features $r^{+} = \sum_{a_i > 0} a_i$ and $r^{-} = |\sum_{a_i < 0} a_i|$ are computed, normalized by $1/10$, and projected through a 2-layer MLP to $d_{\text{emb}}/2$. The $(N_p + N_s)$ position embeddings and the weight embedding are concatenated and projected to $d_{\text{emb}}$ via a final MLP.

\textbf{CoefficientEncoder.}
Coefficients are elements of $\mathbb{Z}_p$ with $p = 2^{31} - 1$. Each coefficient is first mapped to a signed representative in $(-p/2, p/2]$.  Two parallel encodings are computed: (i)~a \emph{small-value} embedding via a learned lookup table for coefficients in $[-31, 31]$, producing a vector of dimension $d_{\text{emb}}/2$; and (ii)~a \emph{large-value} feature vector consisting of $\log(1 + |c|)/20$, $\text{sign}(c)$, $|c| \bmod 100 / 100$, and an indicator for $|c| \leq 100$, projected through a 2-layer MLP to $d_{\text{emb}}/2$. The two representations are concatenated and projected to~$d_{\text{emb}}$.

\textbf{SectorEncoder (Bit Embed).}
The $N_p$-bit sector mask is encoded by assigning each bit position its own learned embedding table mapping $\{0,1\} \to \mathbb{R}^{d_{\text{emb}}/N_p + 1}$. The $N_p$ per-bit embeddings are concatenated and projected to $d_{\text{emb}}$ via a 2-layer MLP with GELU activation and layer normalization.

\textbf{ActionEncoder (TemplateEmbed + IntegralEnc).}
Each action $(a, \mathbf{s})$ is encoded by concatenating a template embedding with a seed encoding. The template index $a$ (one of $N_{\text{op}}$ IBP/LI templates) is mapped to $d_{\text{emb}}/2$ via a learned embedding table. The seed $\mathbf{s} \in \mathbb{Z}^{N_p + N_s}$ is encoded to $d_{\text{emb}}/2$ using a separate \texttt{IntegralEncoder} instance. The two halves are concatenated and projected to $d_{\text{emb}}$ via a 2-layer MLP with ReLU and layer normalization.

\textbf{SubstitutionEncoder (Attn Pooling).}
The substitution history consists of up to $M$ substitutions, each mapping a key integral to a sum of replacement terms.  For each substitution, the key integral is encoded via \texttt{IntegralEncoder}($d_{\text{emb}}$), while each replacement term (integral, coefficient) is encoded by concatenating \texttt{IntegralEncoder}($d_{\text{emb}}/2$) and \texttt{CoefficientEncoder}($d_{\text{emb}}/2$) outputs. The variable-length set of replacement-term embeddings is aggregated into a single vector via learned-query attention pooling: a learnable query vector attends to the replacement embeddings using multi-head attention. The key embedding and pooled replacement embedding are concatenated and projected to $d_{\text{emb}}$ via an MLP with GELU activation.

The sequence of per-substitution embeddings is augmented with learned positional encodings and processed by a 2-layer Transformer encoder (4 heads). A final learned-query attention pool produces a single $d_{\text{emb}}$-dimensional substitution history embedding.

\section{Worked Example: One-Step Worker Episode}
\label{app:worked_example}

We illustrate the mechanics of a single one-step worker by tracing the reduction of $I[1,2,1,1,1,1,-3]$ in the top sector (111111). This integral has weight $w = (7, 3)$, and the worker's goal is to reduce the maximum weight by at least one level. The worker starts with a single-term expression and empty substitution history, then applies three actions chosen by the \textsc{Sailir} classifier via beam search. All arithmetic is performed modulo $p = 1009$.

\textbf{Initial state.} The expression is
\begin{equation}
\mathcal{E} = 1 \cdot I[1,2,1,1,1,1,-\!3], \quad w = (7,3).
\end{equation}

\textbf{Step 0.} The target is $I[1,2,1,1,1,1,-\!3]$. The classifier selects the template $\partial/\partial k_2^\mu \cdot p_1^\mu$ with seed $\mathbf{s} = (1,1,1,1,1,1,-\!3)$. Solving for the target:
\begin{align}
I[1,2,1,1,1,1,-\!3] &= 991 \cdot I[1,1,1,1,2,1,-\!3] \notag \\
&\quad + 1 \cdot I[1,1,1,1,1,2,-\!3] \notag \\
&\quad + \text{(8 subsector terms)}.
\label{eq:step0_sol}
\end{align}
After substituting into $\mathcal{E}$, the expression has 10 terms:
\begin{align}
\mathcal{E} &= 991 \cdot I[1,1,1,1,2,1,-\!3] + 1 \cdot I[1,1,1,1,1,2,-\!3] \notag \\
&\quad + \text{(8 subsector terms)}.
\end{align}
Two integrals remain in the top sector, both at $w = (7,3)$, so $w_{\max} = (7,3)$.

\textbf{Step 1.} The new target is $I[1,1,1,1,1,2,-\!3]$ at $w = (7,3)$. The classifier selects $\partial/\partial k_2^\mu \cdot k_2^\mu$ with seed $\mathbf{s} = (1,1,1,1,1,1,-\!3)$. The raw IBP equation includes the target; after applying the one accumulated substitution, the target still appears (a direct action). Solving:
\begin{align}
I[1,1,1,1,1,2,-\!3] &= 962 \cdot I[1,1,1,1,2,1,-\!3] \notag \\
&\quad + 973 \cdot I[1,1,1,1,1,1,-\!3] \notag \\
&\quad + \text{(4 subsector terms)}.
\end{align}
After substituting into $\mathcal{E}$, the expression simplifies to 6 terms:
\begin{align}
\mathcal{E} &= 944 \cdot I[1,1,1,1,2,1,-\!3] + 973 \cdot I[1,1,1,1,1,1,-\!3] \notag \\
&\quad + \text{(4 subsector terms)}.
\end{align}
Two integrals remain in the top sector: one at $w = (7,3)$ and one at $w = (6,3)$.

\textbf{Step 2.} The remaining $w = (7,3)$ target is $I[1,1,1,1,2,1,-\!3]$. The classifier selects $\partial/\partial k_2^\mu \cdot k_1^\mu$ with seed $\mathbf{s} = (1,1,1,1,1,1,-\!2)$. The raw equation again includes the target, which again persists after applying two accumulated substitutions. Solving:
\begin{align}
I[1,1,1,1,2,1,-\!3] &= 1008 \cdot I[1,1,1,1,1,2,-\!2] \notag \\
&\quad + 47 \cdot I[1,1,1,1,2,1,-\!2] \notag \\
&\quad + \text{(10 subsector terms)}.
\end{align}
After substituting into $\mathcal{E}$, the final expression has 17 terms:
\begin{align}
\mathcal{E} &= 981 \cdot I[1,1,1,1,2,1,-\!2] + 65 \cdot I[1,1,1,1,1,2,-\!2] \notag \\
&\quad + 973 \cdot I[1,1,1,1,1,1,-\!3] \notag \\
&\quad + \text{(14 subsector terms)}.
\end{align}
Three integrals remain in the top sector: two at $w = (7,2)$ and one at $w = (6,3)$.  Since $w_{\max}$ decreased from $(7,3)$ to $(7,2)$, the worker reports success and returns the result to the orchestrator. The substitution history (3 entries) is discarded.

The worker has expressed $I[1,2,1,1,1,1,-\!3]$ at $w = (7,3)$ entirely in terms of in-sector integrals at $w \leq (7,2)$ and subsector integrals. The 17 non-master integrals on the right-hand side themselves need further reduction; the orchestrator caches this result and submits new workers for each one that has not already been cached. This cascading structure---where reducing one integral introduces new integrals that themselves require reduction---is precisely what the hierarchical orchestrator manages automatically via memoized job submission.

\begin{table*}
\centering
\begin{tabular}{c l c l l c c c}
\toprule
Step & Target & $w_{\text{target}}$ & Action & Seed & $w_{\max}$ & \multicolumn{2}{c}{Non-masters} \\
 & & & & & & all & sector \\
\midrule
1 & $I[1,0,-\!1,1,2,0,0]$ & $(4,1)$ & $\partial_{k_2}\!\cdot\! p_2$ & $I[1,0,-\!1,1,2,0,0]$ & $\mathbf{(5,2)}$ & 9 & 8 \\
2 & $I[1,0,-\!1,1,3,-\!1,0]$ & $(5,2)$ & $\partial_{k_2}\!\cdot\! k_1$ & $I[1,0,0,1,1,0,0]$ & $(4,1)$ & 4 & 3 \\
3 & $I[1,-\!1,0,1,2,0,0]$ & $(4,1)$ & $\partial_{k_2}\!\cdot\! k_2$ & $I[1,0,0,1,1,0,0]$ & $(4,1)$ & 4 & 3 \\
4 & $I[1,0,0,1,2,0,-\!1]$ & $(4,1)$ & $\partial_{k_1}\!\cdot\! k_1$ & $I[1,0,0,1,2,0,-\!1]$ & $\mathbf{(5,1)}$ & 5 & 3 \\
5 & $I[1,0,0,2,2,0,-\!1]$ & $(5,1)$ & $\partial_{k_1}\!\cdot\! p_1$ & $I[1,0,0,1,2,0,-\!1]$ & $(5,1)$ & 6 & 3 \\
6 & $I[2,0,0,1,2,0,-\!1]$ & $(5,1)$ & $\partial_{k_1}\!\cdot\! p_2$ & $I[1,0,0,1,2,0,0]$ & $(5,0)$ & 9 & 4 \\
7 & $I[2,0,0,1,2,0,0]$ & $(5,0)$ & $\partial_{k_1}\!\cdot\! p_1$ & $I[1,0,0,1,2,0,0]$ & $(5,0)$ & 9 & 3 \\
8 & $I[1,0,0,2,2,0,0]$ & $(5,0)$ & $\partial_{k_1}\!\cdot\! k_1$ & $I[1,0,0,1,2,0,0]$ & $(4,0)$ & 8 & 2 \\
\bottomrule
\end{tabular}
\caption{Weight trajectory of a non-monotonic 8-step \textsc{Sailir} episode reducing $I[1,0,-\!1,1,2,0,0]$ in sector $(1,0,0,1,1,0)$, starting at $w = (4,1)$. Each step applies the IBP/LI identity specified by the Action (operator) evaluated at the Seed, then solves for the Target. $w_{\text{target}}$: weight of the integral being eliminated. $w_{\max}$: maximum weight among non-master integrals in the target sector after the step. The last two columns count non-master integrals remaining in the expression: ``all'' across all sectors, and ``sector'' restricted to the target sector. Bold $w_{\max}$ entries mark weight increases.}
\label{tab:nonmonotonic}
\end{table*}

\section{Example of a Non-Monotonic   Reduction Path}
\label{app:nonmonotonic}

A crucial feature of IBP reduction is that valid actions are \emph{not} guaranteed to reduce weight---the vast majority \emph{increase} it. In this appendix we provide a detailed example of an integral that is reduced by \textsc{Sailir} along a non-monotonic reduction path.

We trace the reduction of $I[1,0,-\!1,1,2,0,0]$ in sector $(1,0,0,1,1,0)$, starting at weight $w = (4,1)$.  Table~\ref{tab:nonmonotonic} summarizes the 8-step \textsc{Sailir} trajectory.  We see that the maximum weight of the expression fluctuates up and down during the trajectory before the last non-master integral in the sector is eliminated.

\begin{table}[t!]
\centering
\begin{tabular}{r c r r r}
\toprule
Step & $w_{\max}$ & Valid & \multicolumn{2}{c}{Non-masters} \\
 & & actions & all & sector \\
\midrule
1 & $(4,1)$ & 56 & 4 & 3 \\
8 & $(4,2)$ & 296 & 12 & 8 \\
\hline
9 & $\mathbf{(4,3)}$ & 330 & 13 & 9 \\
20 & $(4,3)$ & 758 & 26 & 16 \\
\hline
21 & $\mathbf{(4,4)}$ & 782 & 29 & 19 \\
41 & $(4,4)$ & 1478 & 68 & 35 \\
\hline
42 & $\mathbf{(4,5)}$ & 1512 & 69 & 36 \\
50 & $(4,5)$ & 1823 & 31 & 21 \\
\bottomrule
\end{tabular}
\caption{The naive policy of  minimizing $w_{\max}$ applied to $I[1,0,-\!1,1,2,0,0]$.  Rows show the first and last step of each $w_{\max}$ plateau.  Bold entries mark forced escalations.  After 50 steps, both the maximum weight and the number of sector non-master integrals have \emph{grown} significantly.}
\label{tab:naive_wmax}
\end{table}

\begin{table}[t!]
\centering
\begin{tabular}{r c r r r}
\toprule
Step & $w_{\max}$ & Valid & \multicolumn{2}{c}{Non-masters} \\
 & & actions & all & sector \\
\midrule
0 & $(5,1)$ & 56 & 4 & 2 \\
2 & $(4,1)$ & 151 & 4 & 3 \\
\hline
10 & $(4,1)$ & 488 & 3 & 3 \\
11 & $(3,0)$ & 556 & 1 & 1 \\
\hline
14 & $\mathbf{(5,0)}$ & 644 & 3 & 1 \\
17 & $\mathbf{(6,0)}$ & 752 & 4 & 1 \\
21 & $\mathbf{(7,0)}$ & 890 & 5 & 1 \\
26 & $\mathbf{(8,0)}$ & 1062 & 6 & 1 \\
32 & $\mathbf{(9,0)}$ & 1264 & 7 & 1 \\
39 & $\mathbf{(10,0)}$ & 1500 & 8 & 1 \\
47 & $\mathbf{(11,0)}$ & 1766 & 9 & 1 \\
49 & $(11,0)$ & 1841 & 11 & 1 \\
\bottomrule
\end{tabular}
\caption{Alternative naive policy (minimize sector non-masters) applied to $I[1,0,-\!1,1,2,0,0]$.  The sector count stays at 1 from Step~11 onward, but $w_{\max}$ escalates without bound: the naive policy repeatedly applies $\partial_{k_1}\!\cdot\! k_1$ to raise the highest propagator index by 1, creating an infinite staircase in $r$.}
\label{tab:naive_secnm}
\end{table}

After 8 steps, the episode terminates because the maximum weight of the in-sector integrals has reduced from $(4,1)$ to $(4,0)$. However, we have gone from one non-master integral to 8, with 6 belonging to subsectors of $(1,0,0,1,1,0)$. Nevertheless, the process of reduction, if continued by the orchestrator and the workers, must eventually converge to only master integrals, since the weights provide a total ordering of the integrals.

To illustrate the utility of our trained MDP agent, and to quantify the cost of avoiding weight increases, we simulate a naive policy that also targets the highest-weight non-master integral, but always selects the valid action minimizing $w_{\max}$ of the resulting expression. The (catastrophically bad) result of this naive policy is shown in Table~\ref{tab:naive_wmax}. We see that despite always picking the IBP/LI identity that minimizes the resulting maximum weight, it nevertheless increases steadily, with the total number of terms also growing significantly.
\textsc{Sailir} escapes this trap by accepting temporary weight increases that open reduction pathways.  This is the core capability that beam search with a learned classifier provides: not just avoiding bad moves, but recognizing which superficially bad moves lead to genuinely simpler expressions.

\begin{table*}[t!]
\centering
\begin{tabular}{rr rr rr r rr r}
\toprule
& & \multicolumn{2}{c}{Time (s)} & \multicolumn{2}{c}{Memory (MB)} & & \multicolumn{2}{c}{\textsc{Sailir} Details} & \\
\cmidrule(lr){3-4} \cmidrule(lr){5-6} \cmidrule(lr){8-9}
$r$ & $s$ & \prog{Kira} & \textsc{Sailir} & \prog{Kira} & \textsc{Sailir} & Mem.\ ratio & Jobs & Cache hits & Steps \\
\midrule
10 & 4 & 8.4 & 601 & 159 & 2801 & 17.6 & 3815 & 6142 & 18774 \\
11 & 4 & 14.5 & 658 & 265 & 3016 & 11.4 & 6426 & 11120 & 28780 \\
12 & 4 & 24.2 & 1061 & 418 & 2980 & 7.1 & 7806 & 15497 & 37684 \\
13 & 4 & 36.0 & 1508 & 640 & 2689 & 4.2 & 9096 & 17227 & 41321 \\
\cmidrule(lr){1-10}
10 & 5 & 25.6 & 1318 & 350 & 3212 & 9.2 & 4902 & 9199 & 28316 \\
11 & 5 & 42.4 & 1680 & 500 & 3052 & 6.1 & 8317 & 17220 & 53123 \\
12 & 5 & 69.5 & 1808 & 756 & 3167 & 4.2 & 10390 & 20671 & 57525 \\
13 & 5 & 105.0 & 1719 & 922 & 3251 & 3.5 & 19763 & 47064 & 93300 \\
\cmidrule(lr){1-10}
10 & 6 & 102.8 & 1232 & 1078 & 2813 & 2.6 & 6421 & 11940 & 35536 \\
11 & 6 & 161.4 & 4223 & 1443 & 3132 & 2.2 & 9038 & 20363 & 50977 \\
12 & 6 & 242.0 & 1494 & 1942 & 2975 & 1.5 & 17255 & 39267 & 78851 \\
13 & 6 & 371.3 & 1947 & 2691 & 3325 & 1.2 & 19135 & 45489 & 84782 \\
\cmidrule(lr){1-10}
10 & 7 & 1040 & 4852 & 4645 & 3481 & 0.7 & 11442 & 31552 & 66112 \\
11 & 7 & 1467 & 1496 & 5655 & 3066 & 0.5 & 11292 & 24972 & 60650 \\
12 & 7 & 2086 & 5510 & 6972 & 3612 & 0.5 & 23244 & 62890 & 130598 \\
13 & 7 & 3087 & 3240 & 8723 & 3389 & 0.4 & 26553 & 66576 & 138247 \\
\bottomrule
\end{tabular}
\caption{Benchmark comparison of \prog{Kira} vs.\ \textsc{Sailir} on 16 integrals from the two-loop triangle-box topology. Time: \prog{Kira} total time vs.\ \textsc{Sailir} ideal parallel time (longest chain of sequential dependencies). Memory: \prog{Kira} peak memory vs.\ \textsc{Sailir} maximum per-CPU worker memory. Mem.\ ratio: \textsc{Sailir} memory / \prog{Kira} memory (values $< 1$ indicate \textsc{Sailir} uses less memory). Jobs: total one-step worker jobs submitted. Cache hits: number of times a previously computed reduction was reused. Steps: total beam search steps across all workers.}
\label{tab:benchmark}
\end{table*}

Finally, just for completeness we also illustrate the outcome of an alternative naive policy that always selects the action minimizing the number of sector non-masters (breaking ties by $w_{\max}$). Table~\ref{tab:naive_secnm} shows this policy's trajectory.  It maintains the sector non-master count at 1 from Step~11 onward, but at the cost of unbounded propagator weight escalation. After 50 steps the expression has 11 non-master integrals (1 in the sector) but $w_{\max} = (11,0)$. Contrast this with the \textsc{Sailir} trajectory, which is seen to also be non-monotonic in the number of sector non-master integrals. Clearly, the learned policy is needed for its flexibility and learned intuition, the correct action at any step being impossible to predict from any simple, naive strategy.

\section{Detailed Benchmark Results}
\label{app:benchmark_table}

In Table~\ref{tab:benchmark} we list the detailed comparison of the reduction by \prog{Kira} and \textsc{Sailir}
of the 16 integrals from the two-loop triangle-box topology.

\bibliographystyle{apsrev4-1}
\nocite{apsrev42Control}
\bibliography{refs}

\end{document}